\documentclass[article]{jss}

\usepackage{orcidlink,thumbpdf,lmodern}

\usepackage{framed}

\usepackage{xfrac}

\author{Emma Govan~\orcidlink{0009-0000-9529-6163}\\Hamilton Institute\\Maynooth University \And Andrew L. Jackson~\orcidlink{0000-0001-7334-0434}\\Department of Zoology\\School of Natural Sciences\\Trinity College Dublin \And Richard Inger\\Centre for Ecology\\and Conservation\\University of Exeter \AND Stuart Bearhop\\Centre for Ecology\\and Conservation\\University of Exeter \And Andrew C. Parnell\\Hamilton Institute\\Maynooth University}
\Plainauthor{Emma Govan, Andrew L Jackson, Richard Inger, Stuart Bearhop, Andrew C Parnell}

\title{\pkg{simmr}: A package for fitting Stable Isotope Mixing Models in \proglang{R}}
\Plaintitle{simmr: A package for fitting Stable Isotope Mixing Models in R}
\Shorttitle{A package for SIMMs in \proglang{R}}

\Abstract{
We introduce an \proglang{R} package for fitting Stable Isotope Mixing Models (SIMMs) via both Markov chain Monte Carlo and Variational Bayes. The package is mainly used for estimating dietary contributions from food sources taken via measurements of stable isotope ratios from animals. It can also be used to estimate proportional contributions of a mixture from known sources, for example apportionment of river sediment, amongst many other use cases.  The package contains a simple structure which allows non-expert users to interface with the package, with most of the computational complexity hidden behind the main fitting functions. In this paper we detail the background to these functions and provide case studies on how the package should be used. Further examples are available in the online package vignettes. 
}

\Plainkeywords{ecological modeling, stable isotope mixing models (SIMMs), R, MArkov chain Monte Carlo, Variational Bayes}

\Address{
  Emma Govan\\
  Hamilton Institute \\
  Insight Centre for Data Analytics \\
  Maynooth University\\
  Maynooth\\
  Ireland\\
  E-mail: \email{emma.govan.2021@mumail.ie}\\
}

\begin{document}

\section{Introduction} \label{sec:intro}
Stable Isotope Mixing Models (SIMMs) are a useful tool for ecologists, especially in the reconstruction of animal diets \citep{deniro1978influence}. Starting from stable isotope measurements of animal tissues and their food sources, mixing models allow for estimation of the proportional composition of their food sources in their diet \citep{mckechnie2004stable}. Stable isotope ratios represent the difference in relative abundance of non-radiogenic, stable isotopes expressed as the ratio of an isotope's "heavy" form of an element versus a "light" form relative to an internationally accepted standard. These stable isotope ratios can vary geo-spatially and across the different levels of food webs \citep{hobson1999tracing}. Isotopic data can be obtained relatively easily and allow for many different aspects of diet to be analysed, for example across timescales or locations, depending on the sampled tissues. Typical ecological applications include quantifying animal diets \citep{peterson1987stable}, and estimating the origins of migratory animals \citep{hobson1999tracing}. Whilst isotope ratios were the original data used for these models, other data are used (see later discussion on end member analysis), so these data are referred to more generically as tracers in the \pkg{simmr} package. Similarly, the consumers (usually animals) are referred to more generically as mixtures.

SIMMs require data from both the mixtures being studied as well as all of their sources (for example, the foods they are consuming when we are looking at animal diets). When studying animals, the data can be obtained via tissue samples, such as blood or feathers, depending on the time-frame being studied. For example, isotopes from food are assimilated quickly into tissues with rapid turnover rates such as blood and so this provides a relatively recent estimate of their diet \citep{TieszenTissue1983}; metabolically inactive tissues such as feathers or hair preserve an isotopic record of the diet at the time they were grown \citep{inger2008applications}, and samples from otoliths may provide an overview of the diet of a fish over their lifetime \citep{RadtkeOtolith1996} with successive layers of the tissue being a record of diet through time. Typically, empiricists will make an assumption that a consumer (the mixture) is at equilibrium with its food sources in order to estimate the dietary proportions at a fixed time point. Similarly, users of mixing models are required to assume that all of the potential food sources have been sampled and included in the model \citep{phillips2014best}. A final parameter that needs to be known or estimated is the trophic discrimination factor (or trophic enrichment factor), which describes the change in isotope ratio between the diet and assimilation into consumer proteins. This can be estimated from captive studies of the same species, literature searches of closely related or functionally similar species or in some cases using the software package \pkg{SIDER} \citep{healy2018sider}. 

Our paper covers the maths behind the models used for SIMM analysis in the \proglang{R} \citep{rref} package \pkg{simmr}. We demonstrate how to use the package with an example of Brent Geese data from \cite{inger2006temporal}. The package aims to provide a set of powerful tools, but with a simple to use interface which allows beginners to run models sensibly whilst also allowing advanced users full access to all posterior quantities of the back-end Bayesian model.

The basic mathematical equation for a statistical SIMM is:
$$y = \sum_{k=1}^K p_k s_k + \epsilon$$
where $y$ is the mixture value, $p_k$ are the proportions associated with source $k$ (of $K$ total sources), $s_k$ is the source tracer value for source $k$, and $\epsilon$ is a residual term. In this over-simplification of the model (see Section \ref{sec:maths background} for a more complete version), $s_k$ is an individual level random effect with a given mean and standard deviation, $s_k \sim N(\mu_{sk}, \sigma^2_{sk})$, and the key task of the model is to estimate the $p_k$ proportions. We call $y$ the mixture value for each individual, but they can also be referred to as the consumer value or end member value in the literature.

There are now a number of software tools for fitting SIMMs, which are discussed in detail in Section \ref{sec:software}. Whilst these models are mainly used to study the proportional contribution that different foods make up in an animal's diet, SIMMs can also be used in a wide range of different scenarios. These include the study of a Late Pleistocene bear \citep{mychajliw2020biogeographic}, which confirmed its trophic position is similar to other bears of the same species, and the study of crop usage in Iron Age settlements \citep{styring2022proof}. We expand on this set of examples below. 

Models mathematically identical to SIMMs are also used in many other areas. These are often known as `end member analysis', `mass balance analysis', or `source apportionment'. \cite{hopke1991receptor} is an early review of source apportionment models. It uses linear equations and a least-squares method for running these models. \cite{henry1997history} explores different methods of running these models, such as a geometrical approach. Prior knowledge is then incorporated in \cite{billheimer2001compositional}, which is Bayesian based, and a non-additive error structure is also adopted. \cite{park2001multivariate} incorporates temporal dependence and adopts a MCMC approach to estimate parameters. \cite{lingwall2008dirichlet} uses a Dirichlet prior distribution to allow flexible specification of prior information. The European Union has published several guides on use of source apportionment with receptor models in studies \citep[e.g.][]{european2019, mircea2020european}. 

End member analysis is generally employed by geologists and is used to estimate how different water sources contribute to a mixture. Examples include \cite{soulsby2003identifying}, which employs MCMC methods to study runoff sources during storms in Scotland. \cite{brewer2011source} employs MCMC methods to study runoff sources. Their model allows for consideration of random effects, such as comparison across years. \cite{palmer2008bayesian} operates using MCMC and estimates the proportion that different water sources have contributed to sediment samples. \cite{liu2020quantification} adopts a maximum likelihood method to estimate water sources. Their method employs a multivariate statistical approach to allow for uncertainty in the concentration of end-members, or sources. The end-members contributing to the mixture are first identified and then the proportion that each contributes is calculated. \cite{tao2021endmember} proposes a maximum distance analysis method that estimates both the number and spectral signatures of end-members, which means that it is not essential to know the number and identity of end-members in advance in order for a model to be run.

Another term commonly employed for SIMMs is that of mass balance modelling. The term is often used for apportioning sources of pollution. \cite{christensen2004chemical} evaluates several of these methods, for example weighted least squares and the method of moments. In \cite{campodonico2019chemical} a log-ratio technique is used to analyse how elements move during chemical weathering. \cite{cooper2014sensitivity} uses SIMM-related modelling  to study suspended particulate matter. They look at several different models and provide comparison and advice on choosing an appropriate model. Code is provided for one of their models.   

Our package \pkg{simmr} implements mixing models via both Markov Chain Monte Carlo (MCMC) algorithms and faster Fixed Form Variational Bayes (FFVB). It is not designed to be more fully featured than other \proglang{R} packages that fit SIMMs, rather we aim for a simple unified data structure that enables both non-experts and advanced users to access the tools they need. FFVB is introduced as a foundation to enable much faster fitting of SIMMS, where MCMC can be prohibitively slow. The data structure needed for running SIMMs can be complex, using multiple different data frames of different dimensions, but \pkg{simmr} makes it easy to read in the data and subsequently create plots and model output. We use a Snake case naming convention for consistency and follow the `tidyverse' \citep{tidyverseref} style guide. We aim to keep the number of  functions to a minimum and use S3 classes for access to summary and plot commands. Our visualisations are carefully selected for style and colour choices, and easily produced through built-in functions using ggplot2 \citep{ggplot2ref}. We aim to make all our functions easily extendable so that advanced users can create more complicated outputs. 

\begin{figure}[ht]
\centering
\includegraphics[width=1.0\textwidth]{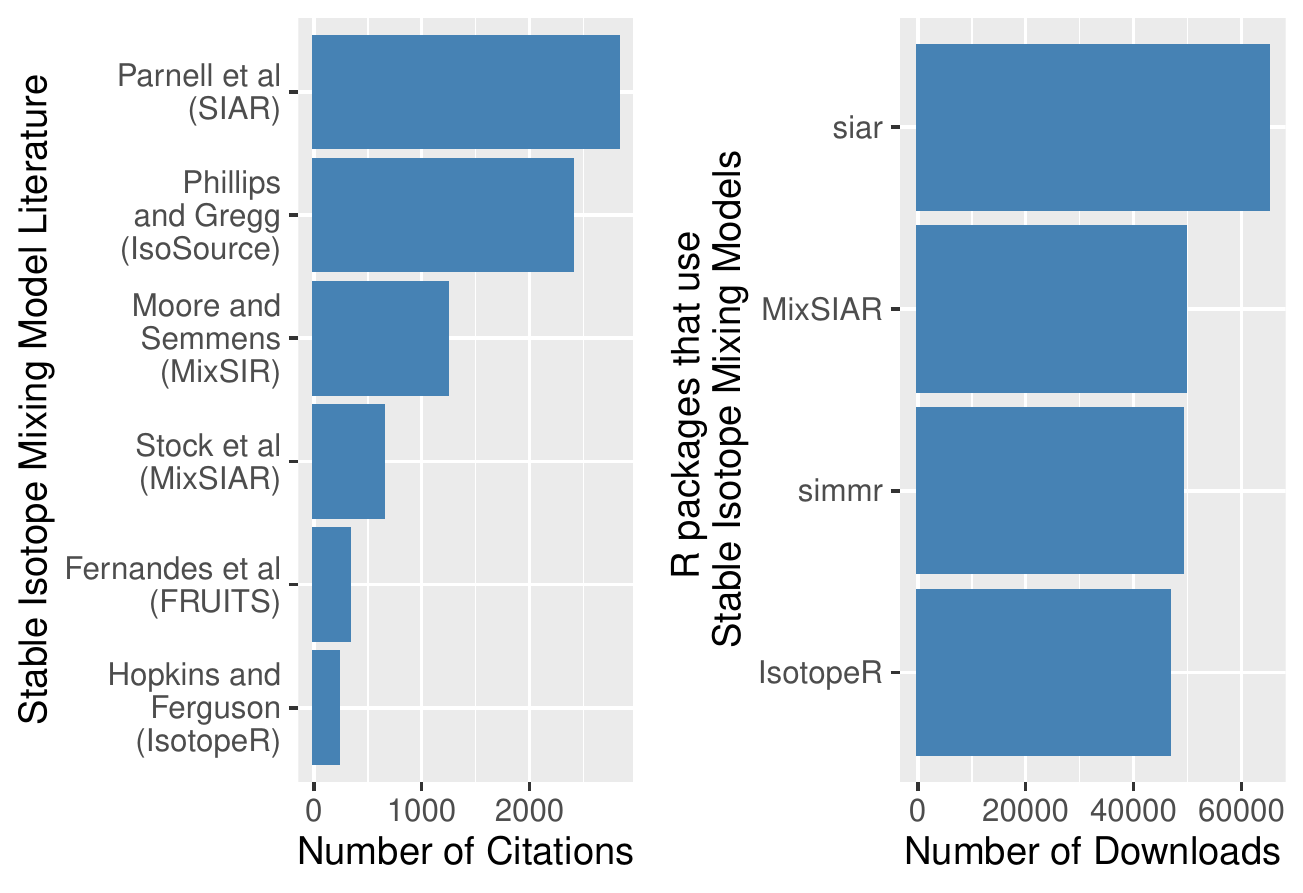}
\caption{\label{fig:downloads_r_packages} Barplot on the left showing the number of downloads that packages using stable isotope mixing models have obtained (download numbers obtained via the cranlogs package \citep{cranlogsref}) and barplot on the right showing the number of citations papers describing SIMMs software have received on Google Scholar (figures correct  as of 9th June 2023).}
\end{figure}

SIMMs are widely used and applicable in many areas. Thus the \proglang{R} packages for running them are frequently downloaded and have received thousands of citations between them. Figure \ref{fig:downloads_r_packages} shows the citation rates of the main papers used for SIMMs and the number of downloads of the associated \proglang{R} packages. 

\section[A short guide to fitting SIMMs using simmr]{A short guide to fitting SIMMs using \pkg{simmr}} \label{sec:how to use}
The first step in using \pkg{simmr} is to install and load the package: 
\begin{Code}
R> install.packages("simmr")
R> library(simmr)
\end{Code}
\pkg{simmr} requires the user to provide mixture data ($y$), source means ($\mu_{sk}$), and source standard deviations ($\sigma_{sk}$). Other variables can be included as described in Section \ref{sec:maths background}.  The data is read into \proglang{R} using the \texttt{simmr\_load} function.
We will use an artificially  generated dataset for illustration:
\begin{CodeChunk}
\begin{CodeInput}
R> y = data.frame(iso1 = c(4, 4.5, 5, 7, 6, 2, 3, 3.5, 5.5, 6.5))
R> mu_s = matrix(c(-10, 0, 10), ncol = 1, nrow = 3)
R> sigma_s = matrix(c(1, 1, 1), ncol = 1, nrow = 3)
R> s_names = c("A", "B", "C")
\end{CodeInput}
\begin{CodeOutput}
\end{CodeOutput}
\end{CodeChunk}
These artificially simple data have measurements of one isotope ratio, named `iso1` and three food sources, labelled A, B, and C. Loading the data into \pkg{simmr} creates an object of class \texttt{simmr\_in}:
\begin{Code}
R> simmr_in_1 = simmr_load(
+       mixtures = y,
+       source_names = s_names,
+       source_means = mu_s,
+       source_sds = sigma_s) 
\end{Code}
We recommend that the user plot these data on an `iso-space' plot before running any model. The iso-space plot shows the isotope(s) ratio on the x (and potentially y) axes. In this case, with one isotope, we need to check that the mixture values lie between the two most extreme values of the food sources on the iso-space plot for the mathematical model to give a reasonable fit to the data. With two isotopes its important to check that the mixture lies within the polygon that can be drawn by joining the food sources with straight lines. The shape created by joining the food sources is referred to as the mixing polygon. The iso-space plot can be generated by running the code:
\begin{CodeChunk}
\begin{CodeInput}
R> plot(simmr_in_1)
\end{CodeInput}
\end{CodeChunk}

\begin{figure}[ht]
\centering
\includegraphics[width=1.0\textwidth]{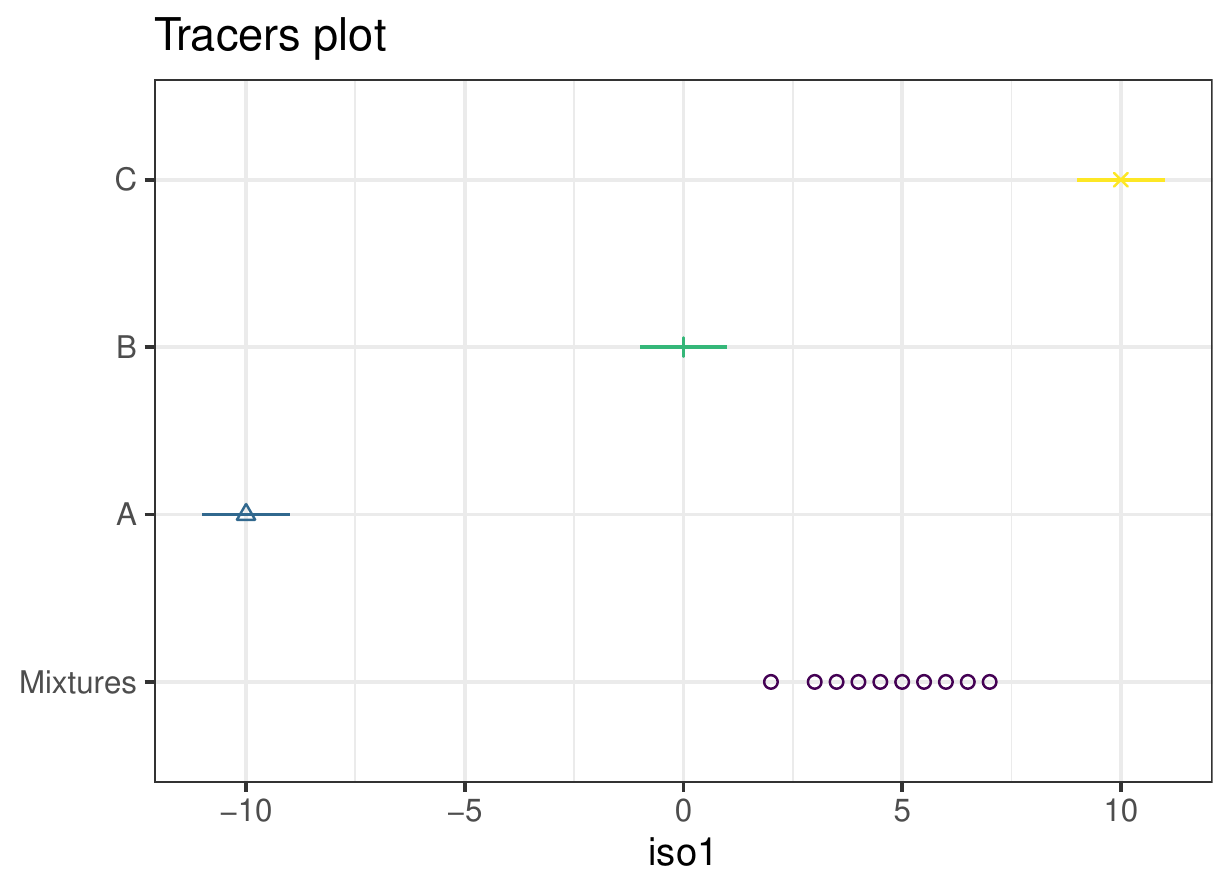}
\caption{\label{fig:simple_simm_1_iso} Simple iso-space plot produced by \pkg{simmr}. The isotope ratios are presented on the x-axis. A, B, and C represent different food sources (with error bars included) and the purple dots represent the mixture values.}
\end{figure}

Figure \ref{fig:simple_simm_1_iso} shows the isospace plot for this simple example, which shows the mixtures lie within the values of the most extreme food sources.The SIMM can then be fitted via MCMC using the \texttt{simmr\_mcmc} function, which produces an object of class \texttt{simmer\_output} and \texttt{mcmc}:

\begin{Code}
R> simmr_out_1 = simmr_mcmc(simmr_in_1)
\end{Code}

When running \texttt{simmr\_mcmc} the first step after running the model is to check convergence. This can be performed by running the following code:
\begin{CodeChunk}
\begin{CodeInput}
R> summary(simmr_out_1, type = "diagnostics")
\end{CodeInput}
\begin{CodeOutput}
Summary for 1 
Gelman diagnostics - these values should all be close to 1.
If not, try a longer run of simmr_mcmc.
deviance        A        B        C sd[iso1] 
    1.00     1.00     1.00     1.01     1.00 
\end{CodeOutput}
\end{CodeChunk}
The values in the diagnostics should all be close to 1. 

\pkg{simmr} produces both textual and graphical summaries of the model run. Starting with the textual summaries, we can get tables of the means, standard deviations and credible intervals (the Bayesian equivalent of a confidence interval) with:
\begin{CodeChunk}
\begin{CodeInput}
R> summary(simmr_out_1, type = "statistics")
\end{CodeInput}
\begin{CodeOutput}
Summary for 1 
           mean    sd
deviance 39.269 2.539
A         0.147 0.065
B         0.243 0.131
C         0.610 0.079
sd[iso1]  1.717 0.602
\end{CodeOutput}
\end{CodeChunk}

This summary provides the mean and standard deviation estimates for the proportion of each food (A, B, and C) that these individuals are eating. It also provides an estimate of the marginal residual error of the isotope ratio (iso1 in this case). Here we can see food C is estimated to make up approximately 61\% of these animals' diet. This finding matches the isospace plot where the consumer isotope ratios are closest to the values for source C.

\pkg{simmr} has built-in functions to allow for visualisation of the results of these models once they have been run. There are multiple options for plotting the output, but perhaps the most useful is the matrix plot: 

\begin{Code}
R> plot(simmr_out_1, type = "matrix")
\end{Code}

\begin{figure}[ht]
\centering
\includegraphics[width=1.0\textwidth]{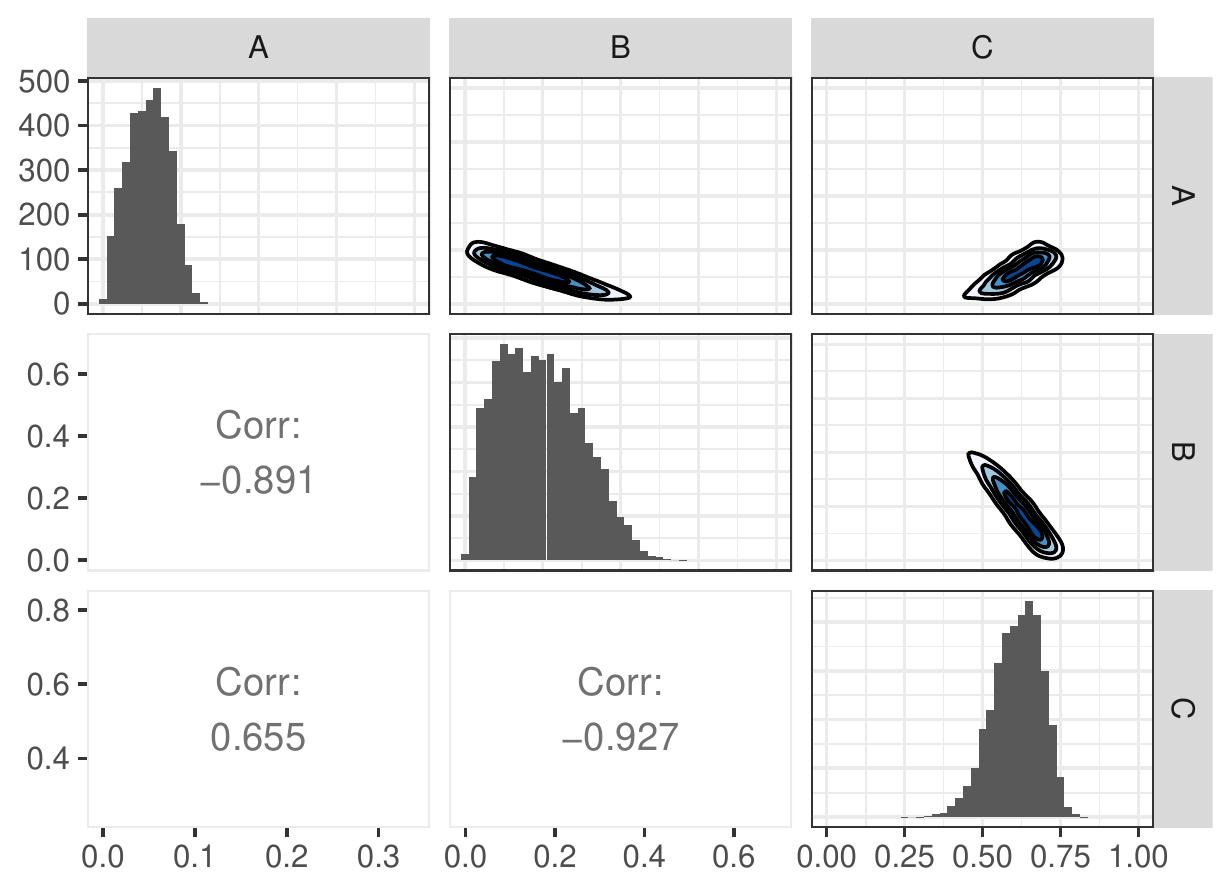}
\caption{\label{fig:mcmc_mat_1_iso} Matrix plot generated from Markov chain Monte Carlo run on Stable Isotope Mixing Model with 1 isotope. Histograms are presented along the diagonal showing the estimated proportion of each food source (A, B, and C) consumed by the individual. The upper-right section shows contour plots. The lower-left section gives correlation values.}
\end{figure}

Figure \ref{fig:mcmc_mat_1_iso} shows histograms of the posterior distribution for the dietary proportions of each source on the diagonal, contour plots of the posterior relationship between the dietary proportions of each food source on the upper-right portion of the plot, and the posterior correlation between the sources on the lower-left portion of the plot. Large negative correlations indicate that the model cannot discern between the two sources; for example they may lie close together in iso-space. Large positive correlations are also possible when there are multiple competing sources. In general, high correlations (negative or positive) are indicative of the model being unable to determine which food sources are being consumed, though the marginal standard deviations can still be narrow. In this case the large negative correlations exist because there is only a single isotope and the model cannot discern, for example, which of sources A and B are pulling the mixture values to the left of source C. 

\section[software]{Other software for fitting stable isotope mixing models} \label{sec:software}

There have been many different software tools developed for the study of SIMMs. Table \ref{tab:package_overview} gives a summary of these tools with a short description. The methods behind these tools includes both Frequentist and Bayesian approaches, and a variety of different fitting techniques. For reference, we provide code to fit our case study data (Section \ref{sec:case study}) using these packages at \url{https://github.com/emmagovan/simmr_paper_SIMM_package_scripts}. 

\begin{table}[ht]
\centering
\begin{tabular}{l l p{4cm} p{7.4cm}}
\hline
Package & Language & Reference & Description \\ \hline
\pkg{Isosource} & Visual Basic & \cite{phillips2003source} & Calculates all possible source combinations and returns feasible solutions\\\hline
\pkg{MixSIR} & Matlab &\cite{moore2008incorporating} & Bayesian - Uses a sampling-importance-resampling algorithm \\\hline
\pkg{siar} & \proglang{R} &\cite{parnell2010source} & Bayesian - uses Markov chain Monte Carlo (MCMC) for its fitting algorithm \\\hline
\pkg{FRUITS} & Visual Basic &\cite{fruits} & Bayesian - allows for consideration of dietary routing. Operates based on Markov chain Monte Carlo simulations. \\\hline
\pkg{MixSIAR} & \proglang{R} &\cite{stock2018analyzing} & Bayesian - allows for consideration of fixed and random effects \\\hline
\pkg{simmr} & \proglang{R} &This paper & Bayesian - can use either MCMC or fixed form variational Bayes (FFVB) \\ \hline
\end{tabular}
\caption{\label{tab:package_overview}} Overview of different software for running Stable Isotope Mixing Models
\end{table}

The first widely available software for fitting SIMMs was \pkg{IsoSource} \citep{phillips2003source}, which worked by generating all possible combinations of dietary proportions that would add to give the isotopic value of the individuals being studied, and presenting all possible combinations to the user. The recommendation was to report the distribution of solutions to avoid any misinterpretation of results. \pkg{IsoSource} was implemented in Visual Basic via the IsoSource computer programme. Whilst \pkg{IsoSource} did not account for many of the intricacies of SIMMs, it was based on several previously developed ideas including concentration dependence, which provides the proportion of each element directly in the food source, \citep[\pkg{IsoConc};][]{phillips2002incorporating} and residual error \citep[\pkg{IsoError};][]{phillips2001uncertainty}.

\pkg{MixSIR} \citep{moore2008incorporating} was later developed and was the first to use a Bayesian framework, based on importance resampling, to estimate dietary proportions. \pkg{MixSIR} works by generating many vectors of possible proportional source contribution and calculating importance weights to determine the posterior distribution. \pkg{MixSIR} is implemented in MATLAB. It allows for several extensions over \pkg{IsoSource} including the ability to account for uncertainty, and incorporation of prior information. See Section \ref{sec:maths background} for a full description of these terms. 

\pkg{SIAR} \citep{parnell2010source} used a Bayesian framework but with Markov chain Monte Carlo (MCMC) as the fitting algorithm. It includes a residual error term and many of the extensions included in MixSIR. The \proglang{R} package is now defunct but maintained on GitHub for backwards compatibility. We have designed \pkg{simmr} to be a replacement for the SIAR software. 

The \pkg{FRUITS} model \citep{fruits} further extended the work by allowing the user to account for the concentration of different food fractions within each food source. \pkg{FRUITS} can also account for a diet-to-tissue signal offset which accounts for different tissues containing different ratios of isotopes, which is the same as concentration dependence discussed in Section \ref{sec:maths background}.
It also simplifies the incorporation of prior information. \pkg{FRUITS} is implemented in Visual Basic via the FRUITS computer programme.

The most recent and perhaps most powerful \proglang{R} package, \pkg{MixSIAR} \citep{stock2018analyzing}, is Bayesian, and allows for consideration of both fixed and random effect covariates on the dietary proportions amongst other extensions. \pkg{MixSIAR} works by creating a custom JAGS \citep[Just Another Gibbs Sampler,][]{plummer2003jags} file for each model run. The package has a number of example data sets included and produces a wide array of output plots and summary statistics. However the model may not be appropriate for novice users and is very slow for complex data sets; which provides the motivation for development of \pkg{simmr} and the incorporation of FFVB.

\section[Maths]{Mathematical background of mixing models} \label{sec:maths background}
The full model implemented in \pkg{simmr} for fitting a SIMM is:
$$y_{ij} \sim N \left( \frac{\sum_{k=1}^K p_k q_{jk} (\mu_{s,jk}+\mu_{c,jk})}{\sum_{k=1}^K p_k q_{jk}}, \frac{\sum_{k=1}^K p_k^2 q_{jk}^2 (\sigma_{s,jk}^2+\sigma_{c,jk}^2)}{(\sum_{k=1}^K p_k q_{jk})^2} +  \sigma_j^2\right)$$
where: 
\begin{itemize}
    \item $y_{ij}$ are the mixture values for individual $i$ on tracer $j$,
    \item $\mu_{s,jk}$ and $\sigma_{s,jk}$ are the mean and standard deviation of the source values for source $k$ on tracer $j$,
    \item $\mu_{c,jk}$ and $\sigma_{c,jk}$ are the mean and standard deviation of the trophic discrimination factors(TFFs or "corrections") for source $k$ on tracer $j$,
    \item $q_{jk}$ represents concentration dependence for tracer $j$ on source $k$,
    \item $p_k$ are the proportions of each source $k$ contributing to the mixture value
    \item $\sigma_j$ is the residual standard deviation on tracer $j$.  
\end{itemize}
The values $y$, $\mu_{s,jk}$, $\sigma_{s,jk}$, $q_{jk}$, $\mu_{c,jk}$, and $\sigma_{c,jk}$ are all given to the model as data. 

The key extension of this model compared to the simple example given in Section \ref{sec:how to use} are that the model now includes multiple tracers, and corrects the proportions for Trophic Discrimination Factors (TDFs) and concentration dependence. As outlined above, TDFs account for the differential loss of one isotope over the other during the assimilation of diet into consumer proteins \citep{inger2008applications}. Different tissues have different macronutrient compositions so TDFs can vary by tissue within a single consumer. Likewise different dietary items contain different elemental proportions (e.g. fats and carbohydrates contain little or no nitrogen when compared to proteins) and a concentration dependence correction can account for this \citep{phillips2002incorporating}. The standard model assumes that a source contributes both elements (in the case of 2 isotopes) equally. Thus a concentration dependence value provides the proportion of that element directly in the food source \citep{phillips2002incorporating}.

As before, the goal of the model fit is to estimate the posterior distribution of $p$ given the data. As we fit the model using the Bayesian paradigm, prior distributions are required for the parameters. The prior for $p$ follows a CLR distribution \citep{clrdistref}:
$$[p_1,...p_k] = \left[ \frac{\exp(f_1)}{\sum_j {\exp(f_j)}}, \ldots, \frac{\exp(f_K)}{\sum_j {\exp(f_j)}} \right]$$
$f$ is then given a multivariate normal distribution:
$$f \sim MVN(\mu_0, \Sigma_0)$$
The values of $\mu_0$ and $\Sigma_0$ can then be set as vague (the defaults are $\mu_0 = 0$ and $\Sigma_0 = \mathbf{I}$) or tuned for informative prior situations (see later functions \texttt{prior\_viz} and \texttt{simmr\_elicit}). The prior distribution on $\sigma$ is set as vague and gamma:
$$\sigma \sim Ga(a, b)$$
where $a$ and $b$ are small values. 
If correction values and concentration are to be used, they must also be provided by the user though they are not necessary to run the model (note: these should be applied in study of animal diets and migration, where TDFs in particular are needed in order to make appropriate inferences). Once the model is run it will then provide posterior samples for $p$, the proportion of each source in the mixture (for example the proportion of each food in the animals diet). Posterior distributions are also available for the parameters $\sigma_j$ which, although are not of primary interest, can also provide some guidance as to the quality of the model fit since they quantify the size of the residual error. 

\section[Fitting SIMMs]{Fitting SIMMs} \label{sec:fitting simms}
\subsection{Fitting using MCMC}
The \texttt{simmr\_mcmc} function allows the user to run their data through a mixing model coded using JAGS. The function has preset general priors for $p$, which can be altered by the user if they wish. The number of chains, iterations, burn-in period, and thinning can also be edited by the user, and are set to sensible default values otherwise. 

The JAGS code for this model is provided as a model string inside the \proglang{R} function. The parameters saved when this model is run are $p$ and $\sigma$. The output is assigned the class \texttt{simmr\_output}. This allows for the package to use one plot function to plot inputs and outputs from both MCMC and FFVB. The function will pick out groups and run a separate MCMC algorithm for each one if needed. These groups can be represented by any categorical variable provided as part of the data. Grouping structures might include: demographic divisions such as age or sex, the same animals measured at different times of year; different packs within the same species; or populations of the same species living in different habitat types.

Often SIMMs need to be run on just a single consumer isotope observation, in which case the residual term becomes unnecessary. In cases where only a single observation is provided to \texttt{simmr\_load} the model uses a prior for $\sigma_j$ with high prior mass on zero. This is termed a `simmr solo` run. All the output plots and summaries work on this structure exactly as they do on a standard \pkg{simmr} run. 

\subsection{Fitting using VB}
The \texttt{simmr\_ffvb} function can be used if the user wishes to fit a SIMM using Fixed Form Variational Bayes (FFVB). FFVB works by approximating the full posterior using a simpler distribution \citep{pati2018statistical}. As it is an optimisation routine, it has the potential to run much faster than MCMC which relies on random sampling. FFVB aims to minimise Kullback-Leibler \citep[KL;][]{kullback1951information}  divergence between the posterior and the VB approximation. We provide a more detailed description of the FFVB fitting approach we use in Appendix \ref{app:algorithm}.

Whilst the fitting method for FFVB is fundamentally different to the MCMC approach, the code still produces posterior samples of $p$ and $\sigma$, and the output is assigned the class \texttt{simmr\_output} as above. The user should not notice any difference in fitting using the two approaches, though fitting complicated models using FFVB should be faster than MCMC.

\section[Brent Geese]{Case study: Brent Geese} \label{sec:case study}
This section provides code and explanation for running a two-isotope model in \pkg{simmr} with data in groups, in this case data on geese gathered at different times of year. The dataset is from \cite{inger2006temporal} and is provided as a sample data set within \pkg{simmr}. To begin we load in the package:
\begin{Code}
R> library(simmr)
\end{Code}
In this example, our mixture is the geese, the sources are the food the geese eat and the tracers are $\delta^{13}C$ and $\delta^{15}N$. 
\pkg{simmr} requires the user to supply consumer data, the source means, and the source standard deviations. Trophic Discrimination Factors (TDFs) and concentration dependence are included in this example. The data is read into \proglang{R} using the \texttt{simmr\_load} function. \pkg{simmr} has the ability to perform repeated runs on data sets if the data is separated into different groups.  In \pkg{simmr} a separate model run will be performed for each group provided a grouping variable is given to \texttt{simmr\_load}.

\begin{figure}[ht]
\centering
\includegraphics[width=1.0\textwidth]{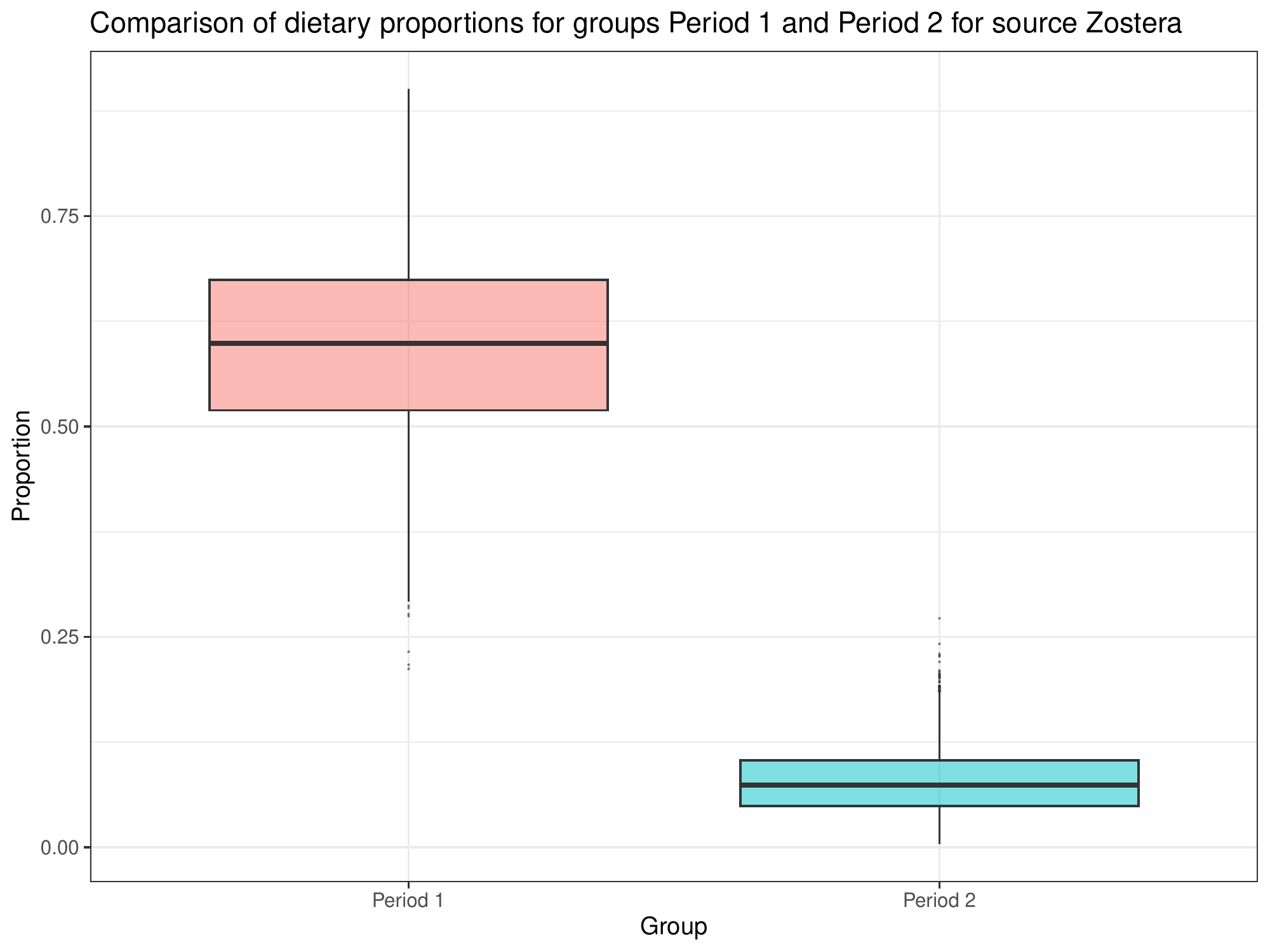}
\caption{\label{fig:compare_groups} Boxplot comparing proportion of Zostera in diet in Period 1 versus Period 2.}
\end{figure}

This data set can be called from the \pkg{simmr} package and an overview of the data can be seen via \texttt{str}:
\begin{CodeChunk}
\begin{CodeInput}
R> str(geese_data)
\end{CodeInput}
\begin{CodeOutput}
List of 9
 $ mixtures           : num [1:251, 1:2] -11.4 -11.9 -10.6 -11.2 -11.7 ...
  ..- attr(*, "dimnames")=List of 2
  .. ..$ : NULL
  .. ..$ : chr [1:2] "d13C_Pl" "d15N_Pl"
 $ tracer_names       : chr [1:2] "d13C" "d15N"
 $ source_names       : chr [1:4] "Zostera" "Grass" "U.lactuca" "Enteromorpha"
 $ source_means       : num [1:4, 1:2] -11.17 -30.88 -11.17 -14.06 6.49 ...
 $ source_sds         : num [1:4, 1:2] 1.21 0.64 1.96 1.17 1.46 2.27 1.11 0.83
 $ correction_sds     : num [1:4, 1:2] 0.63 0.63 0.63 0.63 0.74 0.74 0.74 0.74
 $ concentration_means: num [1:4, 1:2] 0.36 0.4 0.21 0.18 0.03 0.04 0.02 0.01
 $ correction_means   : num [1:4, 1:2] 1.63 1.63 1.63 1.63 3.54 3.54 3.54 3.54
 $ groups             : chr [1:251] "Period 1" "Period 1" "Period 1" "Period 1" ...
\end{CodeOutput}
\end{CodeChunk}
The data can then be used to create an object of class \texttt{simmr\_in}:
\begin{Code}
R> simmr_groups = with(geese_data, 
+       simmr_load(mixtures = mixtures,
+       source_names = source_names,
+       source_means = source_means,
+       source_sds = source_sds,
+       correction_means = correction_means,
+       correction_sds = correction_sds,
+       concentration_means = concentration_means,
+       group = groups))
\end{Code}
We create the recommended iso-space plot  to ensure all the mixtures lie inside the mixing polygon defined by the sources. \texttt{group} specifies which groups we want to plot. \texttt{xlab} and \texttt{ylab} allow for editing of the x and y axes labels. The following code creates an iso-space plot displaying groups 1 to 8. The axes labels are edited here to include the parts-per-mille sign. 
\begin{CodeChunk}
\begin{CodeInput}
R> plot(simmr_groups,
+       group = 1:8,
+       xlab = expression(paste(delta^13, "C (\u2030)", sep = "")), 
+       ylab = expression(paste(delta^15, "N (\u2030)", sep = "")), 
+       title = "Iso-space plot of Inger et al Geese data",
+       mix_name = "Geese")
\end{CodeInput}
\end{CodeChunk}
\begin{figure}[ht]
\centering
\includegraphics[width=1.0\textwidth]{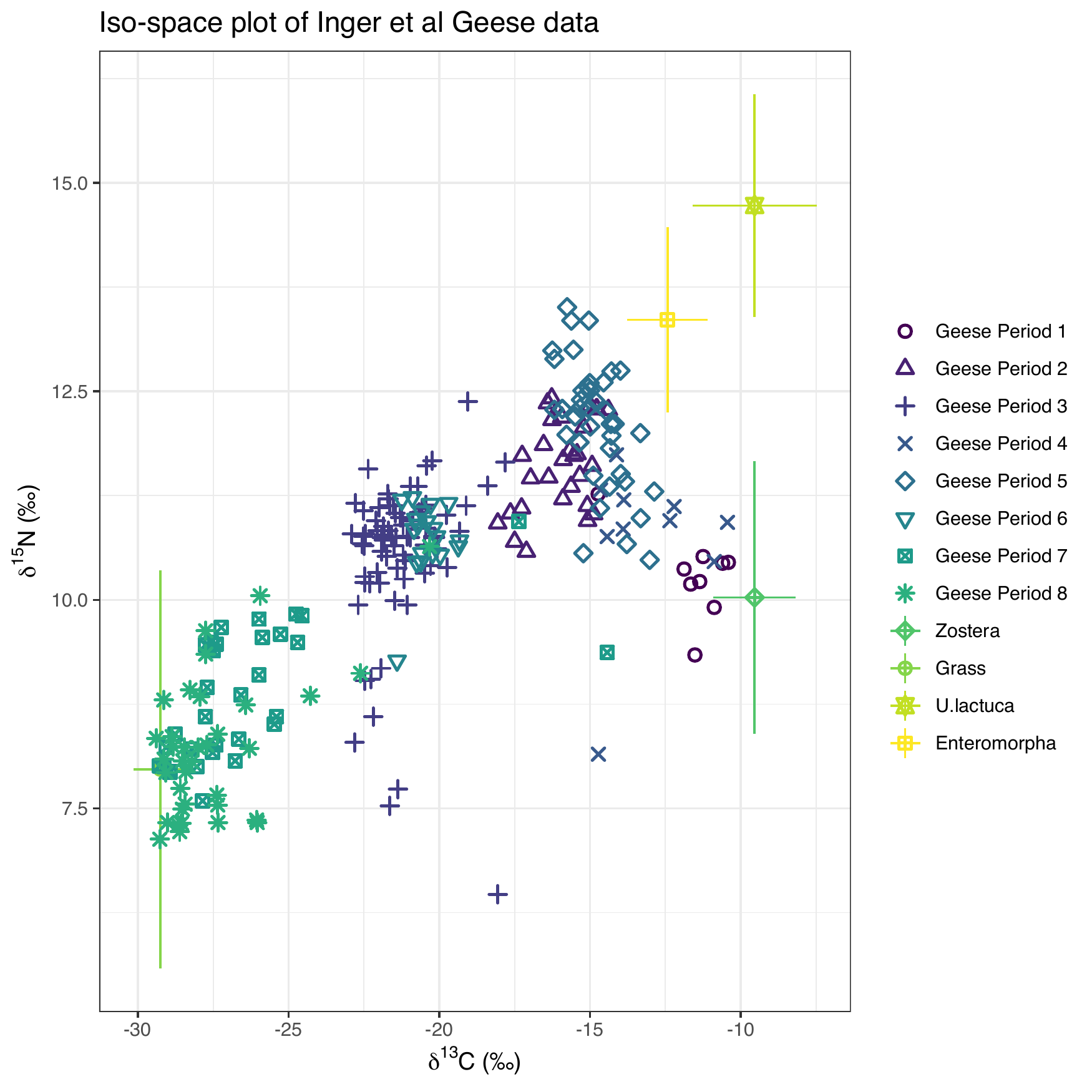}
\caption{\label{fig:group_geese} Iso-space plot of the eight groups of geese as well as their food sources.}
\end{figure}
The isospace plot can be viewed in Figure \ref{fig:group_geese} and it can be seen that all the data lies within the mixing polygon. The SIMM can be run either through JAGS or FFVB. The code is as follows:
\begin{Code}
R> simmr_groups_out = simmr_mcmc(simmr_groups)
R> simmr_groups_out_ffvb = simmr_ffvb(simmr_groups)
\end{Code}
The first step after running \texttt{simmr\_mcmc} is to check convergence, which can be performed by running the code:
\begin{Code}
R> summary(simmr_groups_out, type = "diagnostics")
\end{Code}
It is important that all the diagnostic values are close to 1 - if not, a longer \texttt{simmr\_mcmc} run is recommended.
\pkg{simmr} can produce textual summaries of the model run, an example of which can be seen below. Options within this function include quantiles, statistics, and correlations.
\begin{CodeChunk}
\begin{CodeInput}
R> summary(simmr_groups_out, 
+       type = "quantiles", 
+       group = 1)
\end{CodeInput}
\begin{CodeOutput}
Summary for Period 1 
               2.5%    25%    50%    75%  97.5%
deviance     51.090 52.539 54.047 56.146 62.934
Zostera       0.387  0.521  0.593  0.670  0.812
Grass         0.031  0.058  0.072  0.086  0.119
U.lactuca     0.021  0.071  0.125  0.191  0.321
Enteromorpha  0.024  0.092  0.171  0.278  0.483
sd[d13C_Pl]   0.071  0.523  0.891  1.293  2.332
sd[d15N_Pl]   0.017  0.163  0.356  0.636  1.431
\end{CodeOutput}
\end{CodeChunk}
\pkg{simmr} has built in functions to allow for visualisation of the results of these models once they have been run. Options for plots include matrix plots, boxplots, histograms, and density plots. The code for running a boxplot is as follows:
\begin{Code}
R> plot(simmr_groups_out,
+       type = "boxplot",
+       group = 2,
+       title = "simmr output group 2")
\end{Code}
\begin{figure}[ht]
\centering
\includegraphics[width=1.0\textwidth]{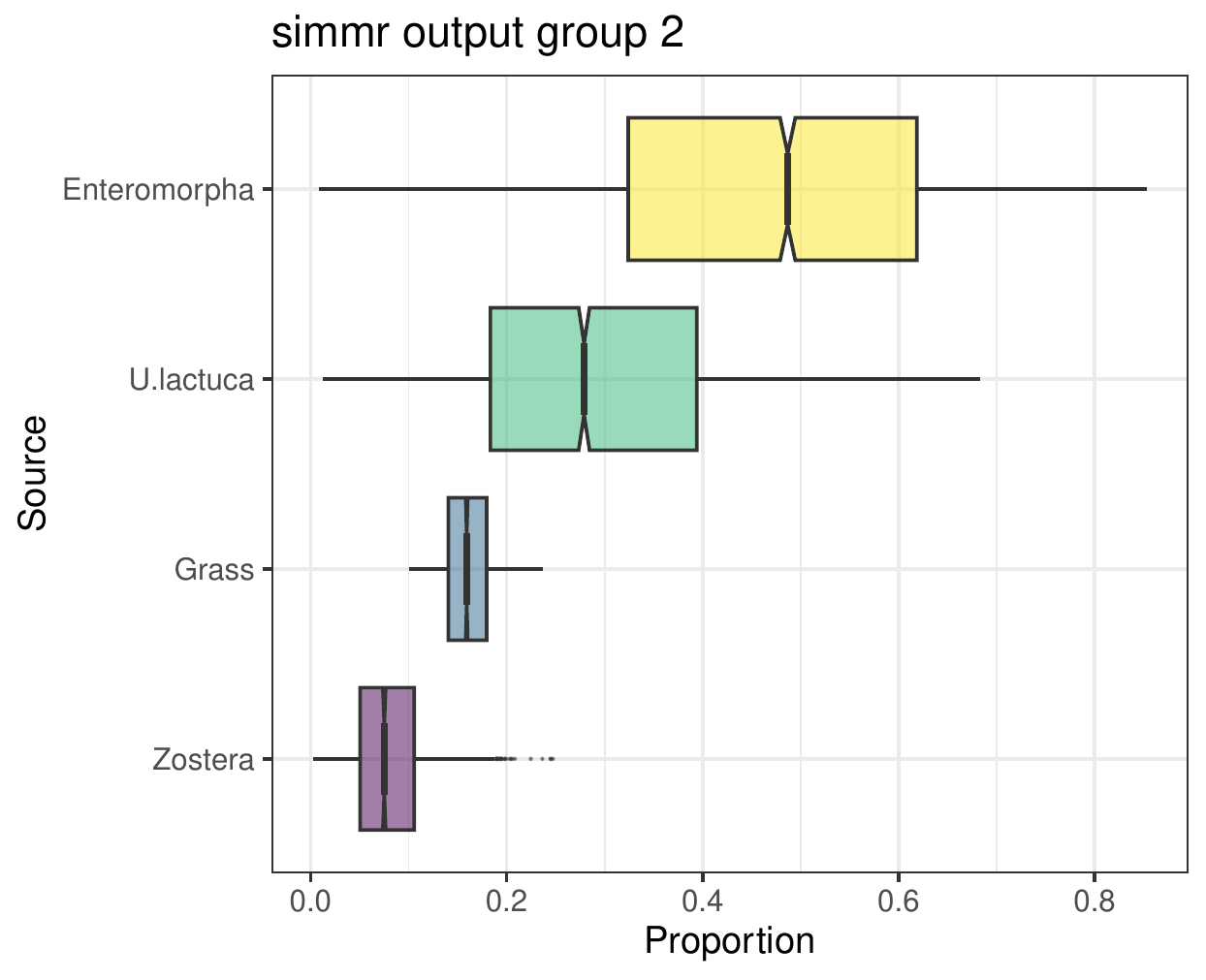}
\caption{\label{fig:mcmc_boxplot} Boxplot of geese group 2 food proportion estimates generated from MCMC run.}
\end{figure}
\begin{Code}
R> plot(simmr_groups_out,
+       type = "matrix",
+       group = 6,
+       title = "simmr output group 6")
\end{Code}
\begin{figure}[ht]
\centering
\includegraphics[width=1.0\textwidth]{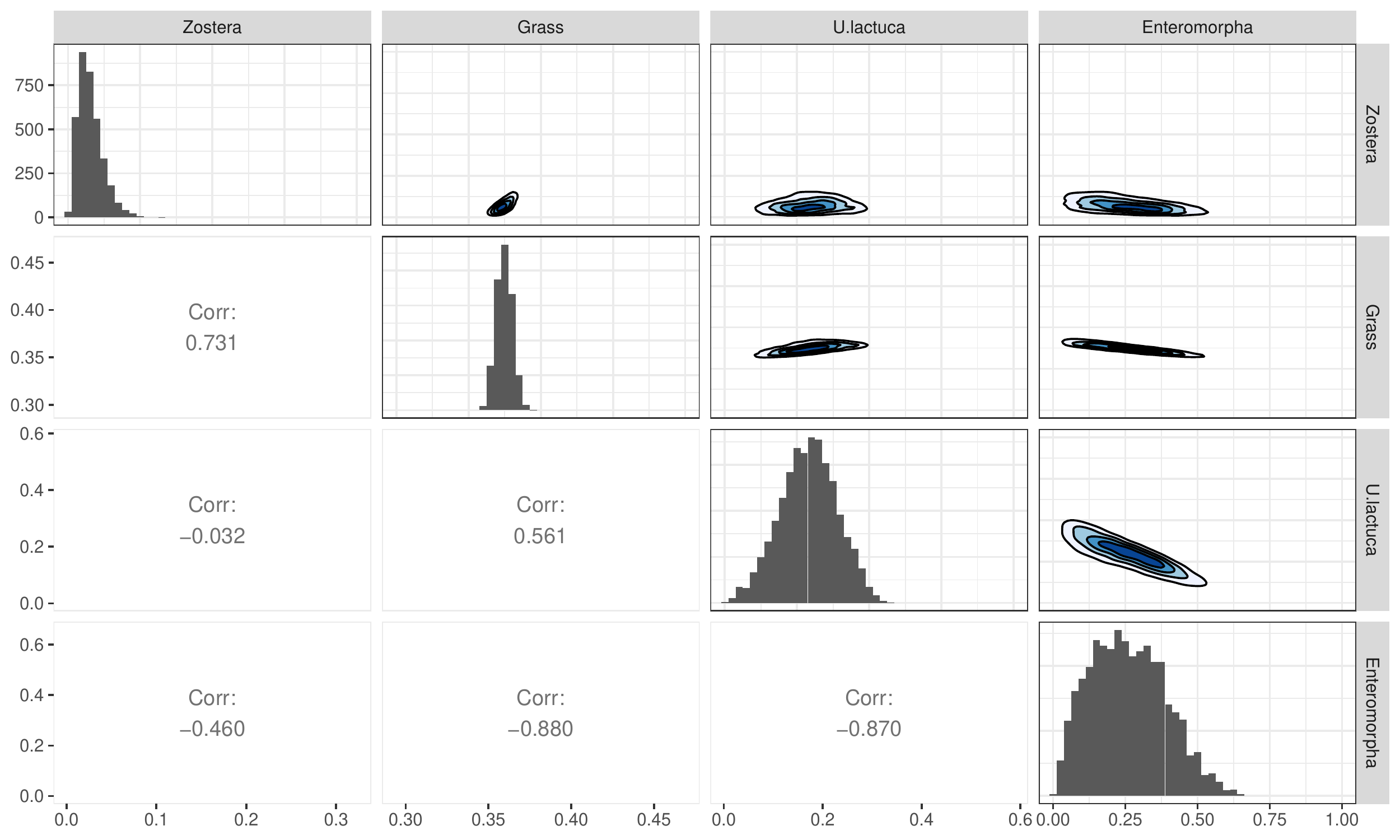}
\caption{\label{fig:mcmc_mat_group_6} Matrix plot of geese group 6 generated from the MCMC run. The diagonal contains histograms showing the estimated proportion of each food. The upper section shows contour plots. The lower section gives correlation values.}
\end{figure}
Figure \ref{fig:mcmc_boxplot} shows boxplots which show the proportion each food source is estimated to make up of the animals diet. The boxplot allows for easy visualisation of the proportions. Figure \ref{fig:mcmc_mat_group_6} shows the histograms of source proportions on the diagonal, contour plots of the relationship between the sources on the upper diagonal, and the correlation between the sources on the lower diagonal. In this case, we can see that the geese are consuming mostly \textit{Enteromorpha spp}, some \textit{Ulva lactuca} and Grass, and hardly any \textit{Zostera spp}.
The \texttt{compare\_groups} and \texttt{compare\_sources} functions in \pkg{simmr} allow for comparison of source consumption across different groups or sources. Below and in Figure \ref{fig:compare_groups} we show the output of comparing the proportions of \textit{Zostera spp} between groups 1 and 2: 
\begin{CodeChunk}
\begin{CodeInput}
R> compare_groups(simmr_groups_out, source = "Zostera", groups = 1:2)
\end{CodeInput}
\begin{CodeOutput}
Prob ( proportion of Zostera in group Period 1 > proportion 
of Zostera in group Period 2 ) = 1
\end{CodeOutput}
\end{CodeChunk}

\begin{figure}[ht]
\centering
\includegraphics[width=1.0\textwidth]{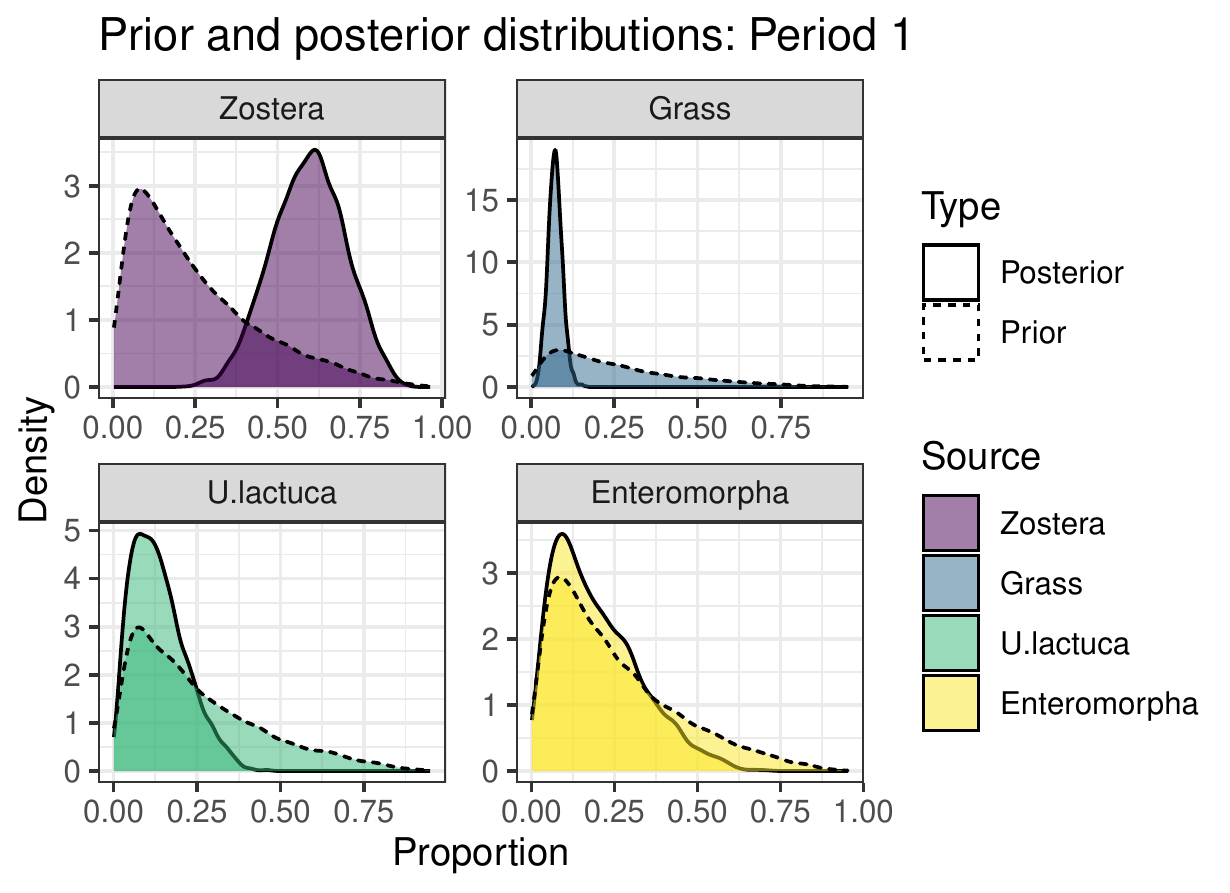}
\caption{\label{fig:prior_viz} Four density plots showing the posterior and prior of each of the four food sources (Zostera, Grass, Ulva lactuca, and Enteromorpha).}
\end{figure}

Beyond the main functions for plotting and summarising SIMMs, \pkg{simmr} contains other functions that may be useful for the user in interpreting output or guiding model development. \texttt{prior\_viz} allows for visualisation of the priors set for the data versus the eventual posterior, and saves the data in a data frame if the user wishes to create their own plots. The function can be especially helpful when some food sources do not provide knowledge about diet and so the posterior dietary proportion reverts to the prior. Figure \ref{fig:prior_viz} shows the results of using the function on the data above with the following code:

\begin{Code}
R> prior <- prior_viz(simmr_groups_out, group = 1)    
\end{Code}

The \texttt{simmr\_elicit} function allows the user to input informative prior information to the model and can be used with both the MCMC and FFVB functions. Prior information about the dietary proportions might come from sources such as direct observation, faeces, stomach contents, or prey remains (\cite{moore2008incorporating}, \cite{franco2013bias}, \cite{hertz2017overwinter}). Finding appropriate prior values for the latent multivariate normal distribution parameters ($\mu_0$ and $\Sigma_0$) can be hard since the prior information is usually available in the dietary proportion space. The function thus runs an optimisation routine to match provided dietary proportions to optimal values of $\mu_0$ and $\Sigma_0$ and provides these to the user so that they can be added as arguments to e.g. \texttt{simmr\_mcmc}. The $\texttt{prior\_viz}$ function can then be used to see the effect of these prior assumptions on the posterior.

When sources lie in similar locations on the iso-space plots it is sometimes desirable to combine sources together. The \pkg{simmr} package allows the user to choose which sources to combine \textit{a posteriori} using the \texttt{combine\_sources} function. The advantage of combining such sources is that the negative covariance between their esimated proportions will reduce the variance of the resulting summed source contribution. The output of \texttt{combine\_sources} is also of class \texttt{simmr\_output} and so can be used with all other plotting and summary functions. 

\begin{figure}[ht]
\centering
\includegraphics[width=1.0\textwidth]{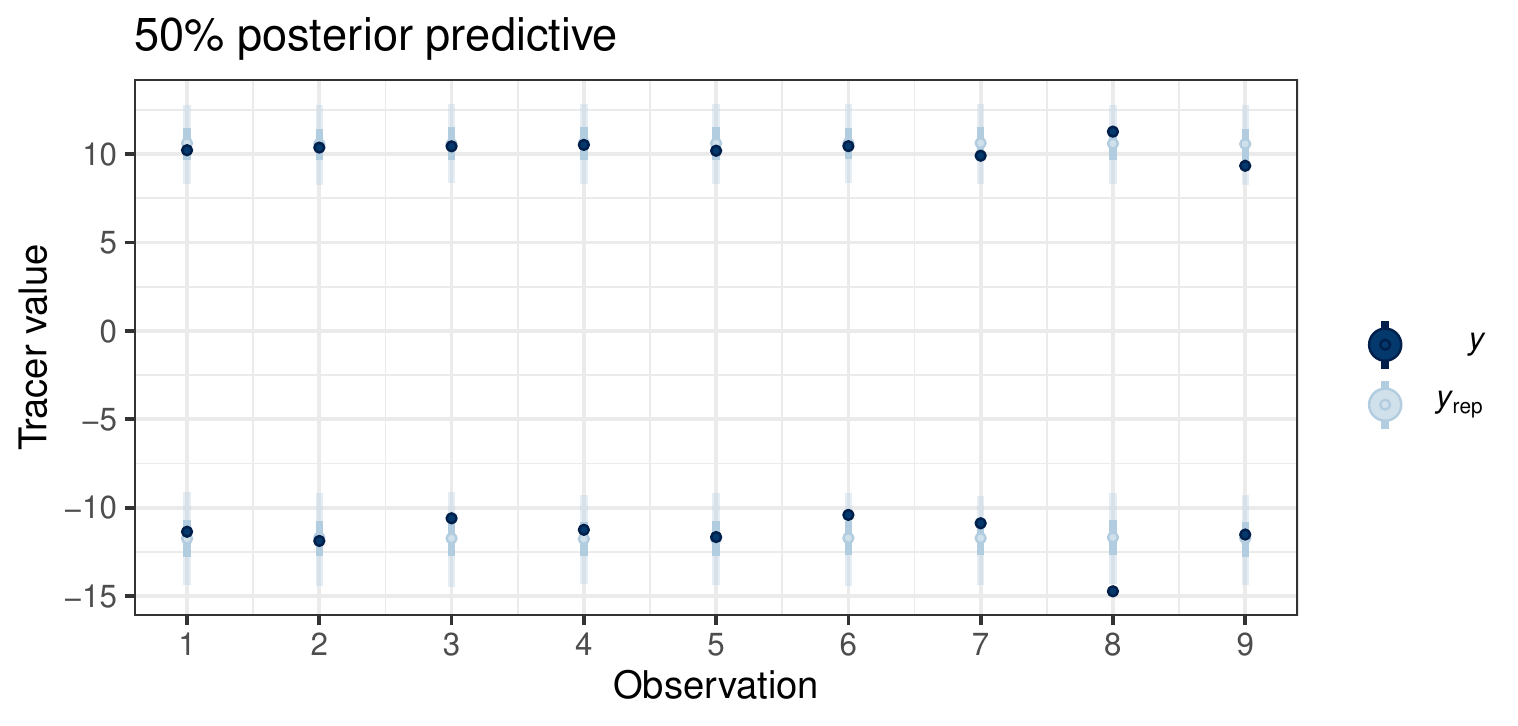}
\caption{\label{fig:post_pred} Posterior predictive distribution of the observations for geese data group one, probability interval = 0.5}
\end{figure}

For a final check of the model fit, the function \texttt{posterior\_predictive} creates the posterior predictive distribution of the observations and plots this for each observation. For models that fit well we would expect, for example, 50\% of observations to be within the 50\% posterior predictive distribution. The function re-runs the JAGS code for the model but with an extra likelihood term inserted to extract the posterior predictive distribution. These values are returned from the function to enable more advanced use of the posterior predictive. The output is seen in Figure \ref{fig:post_pred}. 

\section{Discussion} \label{sec:summary}
Our package, \pkg{simmr} allows analysis of basic SIMMs using either MCMC or FFVB algorithms. It is a convenient tool with built-in class objects that allows analysis to be performed easily. Plots and summaries are simple to produce and the class system makes it easy to process data, even though the underlying data structure can be relatively complex for non-\proglang{R} users.

There are a number of assumptions made which are common to many of these models. Probably the most fundamental is that the consumer is at equilibrium with their food and that their diet is static. In a dynamic system, clearly this assumption is always violated, but careful interpretation of the results can still yield valid insights by shifting the considered time window over which the diet is quantified using SIMMs \citep{Arnoldi2023.04.04.535564}. We assume that we know all the food sources that the animal is eating. If a food source is missing it can affect the shape of the mixing polygon and the resulting diet composition obtained by the model \citep{phillips2014best}. It is highly recommended that the user views the iso-space plot before running the model to ensure that the data points lie within the mixing polygon created by the food sources. If the food sources do not lie within the mixing polygon, it can indicate that a food source has been missed. However, all data lying within the mixing polygon does not guarantee every food source has been found. It is important to consider the biology of the organism, how robust sampling was, and to ensure samples have been obtained of every food the organism eats. Similar assumptions would apply if studying other systems.

There are a number of areas where the user will need to make decisions, for example, whether or not to exclude outliers, and whether food sources should be combined. Combining sources can be done through the \texttt{combine\_sources} function provided by \pkg{simmr} but it is recommended that this is performed \textit{a posteriori}, if it is to be performed, and that there is a sound biological basis for doing so.  

Our new package has the advantage that the models are quick to run, easy to use, and have several built in checks. We have built-in functions which allow customisable high quality plots to be produced that allow for sources to be combined \textit{a posteriori} and for sources and groups to be compared.
Future plans include expansion of this package to allow it to run with more complex models, and include random and fixed effects.

\section*{Computational details}
The results in this paper were obtained using
\proglang{R}~4.2.2. \proglang{R} is available from the Comprehensive
\proglang{R} Archive Network (CRAN) at
\url{https://CRAN.R-project.org/}. \pkg{JAGS} version 4.3.1 was used. 
C++ compiler gcc version 4.2.1.

\section*{Acknowledgments}

\begin{leftbar}
This publication has emanated from research conducted with the financial support of Science Foundation Ireland under Grant number SFI/12/RC/2289\_P2\_Parnell to ACP and Irish Research Council Laureate award IRCLA/2017/186 to ALJ . For the purpose of Open Access, the authors have applied a CC BY public copyright licence to any Author Accepted Manuscript version arising from this submission.
\end{leftbar}

%% -- Bibliography -------------------------------------------------------------
%% - References need to be provided in a .bib BibTeX database.
%% - All references should be made with \cite, \citet, \citep, \citealp etc.
%%   (and never hard-coded). See the FAQ for details.
%% - JSS-specific markup (\proglang, \pkg, \code) should be used in the .bib.
%% - Titles in the .bib should be in title case.
%% - DOIs should be included where available.

\bibliography{refs}

%% -- Appendix (if any) --------------------------------------------------------
%% - After the bibliography with page break.
%% - With proper section titles and _not_ just "Appendix".

\newpage

\begin{appendix}

\section{Fixed Form Variational Bayes Algorithm} \label{app:algorithm}
We use the FFVB algorithm of \citep{tran:2021}. If we define the joint set of parameters as $\theta = (f, \tau)$ where $\tau = \sigma^{-2}$ then we write our factorised variational posterior as:
\begin{eqnarray*}
q_\lambda(\theta) = q(f)q(\tau)
\end{eqnarray*}
where $\lambda = (\mu_f, \Sigma_f, c, d)^T$ is the set of hyper-parameters associated with the variational posteriors:

\begin{eqnarray*}
q(f) \equiv MVN(\mu_f, \Sigma_f) \\
q(\tau) \equiv Ga(c,d)
\end{eqnarray*}

To start the algorithm, initial values are required for $\lambda^{(0)}$ (we use parenthetical super-scripts to denote iterations.), the sample size $S$, the adaptive learning weights ($\beta_1, \beta_2$), the fixed learning rate $\epsilon_0$, the threshold $\alpha$, the rolling window size $t_W$ and the maximum patience $P$. 

Define $h$ to be the log of the joint distribution up to the constant of proportionality:
\begin{eqnarray*}
h(\theta) = \log \left( p(y|\theta) p(\theta) \right)
\end{eqnarray*}

and $h_\lambda$ to be the log of the ratio between the joint and the VB posterior:
\begin{eqnarray*}
h_\lambda(\theta) = \log \left( \frac{ p(y|\theta) p(\theta) }{ q_\lambda(\theta) } \right) = h(\theta) - \log q_\lambda(\theta) 
\end{eqnarray*}

The initialisation stage proceeds with:
\begin{enumerate}
\item Generate samples from $\theta_s \sim q_{\lambda^{(0)}(\theta)}$ for $s=1,...S$

\item Compute the unbiased estimate of the lower bound gradient:
\begin{eqnarray*}
\widehat{\nabla_\lambda{LB}(\lambda^{(0)})} = \frac{1}{S}\sum_{s=1}^S\nabla_\lambda[\log(q_\lambda(\theta_s))] \circ h_\lambda(\theta_s) \bigg\vert_{\lambda = \lambda^{(0)}}
\end{eqnarray*}
where $\circ$ indicates element-wise multiplication

\item Set

\begin{eqnarray*}
\bar{g}_0 := \nabla_\lambda{LB}(\lambda^{(0)})\\
\bar{\nu}_0 := \bar{g}_0^2\\
\bar{g} = g_0\\
\bar{\nu} = \nu_0\\
\end{eqnarray*}

\item Estimate the control variate $c_i$ for the $i$th element of $\lambda$ as:
\begin{eqnarray*}
c_i = \frac{Cov \left(\nabla_{\lambda_i}[\log(q_\lambda(\theta))]h_\lambda(\theta),\nabla_{\lambda_i}[\log(q_\lambda(\theta))]\right)}{ Var(\nabla_{\lambda_i}[\log(q_\lambda(\theta))])}
\end{eqnarray*}

across the samples generated in step 1

\item Set $t=1$, patience = 0, and `stop = FALSE`.\\
\end{enumerate}
Now the algorithm runs with:
\begin{enumerate}
\item Generate samples from $\theta_s \sim q_{\lambda^{(t)}(\theta)}$ for $s=1,...S$

\item Compute the unbiased estimate of the lower bound gradient:
\begin{eqnarray*}
g_t := \widehat{\nabla_\lambda{LB}(\lambda^{(t)})} = \frac{1}{S}\sum_{s=1}^S\nabla_\lambda[\log(q_\lambda(\theta_s))] \circ(h_\lambda(\theta_s) - c) \bigg\vert_{\lambda = \lambda^{(t)}}
\end{eqnarray*}

where $\circ$ indicates element-wise multiplication.

\item Estimate the new control variate $c_i$ for the $i$th element of $\lambda$ as:
\begin{eqnarray*}
c_i = \frac{Cov \left(\nabla_{\lambda_i}[\log(q_\lambda(\theta))]h_\lambda(\theta),\nabla_{\lambda_i}[\log(q_\lambda(\theta))]\right)}{ Var(\nabla_{\lambda_i}[\log(q_\lambda(\theta))])}
\end{eqnarray*}
across the samples generated in step 1

\item Compute: 

\begin{eqnarray*}
v_t = g_t^2 \\
\bar{g} = \beta_1\bar{g} + (1-\beta_1)g_t \\
\bar{v} = \beta_2\bar{v} + (1-\beta_2)v_t \\
\end{eqnarray*}

\item Update the learning rate:
\begin{eqnarray*}
l_t = min(\epsilon_0, \epsilon_0\frac{\alpha}{t})
\end{eqnarray*}
and the variational hyper-parameters:
\begin{eqnarray*}
\lambda^{(t+1)} = \lambda^{(t)} + l_t\frac{\bar{g}}{\sqrt{\bar{v}}}
\end{eqnarray*}

\item Compute the lower bound estimate:
\begin{eqnarray*}
\widehat{LB}(\lambda^{(t)}) := \frac{1}{S} \sum_{s=1}^S h_{\lambda^{(t)}}(\theta_s)
\end{eqnarray*}

\item If $t \ge t_W$ compute the moving average LB
\begin{eqnarray*}
\overline{LB}_{t-t_W+1} := \frac{1}{t_W} \sum_{k=1}^{t_W} \widehat{LB}(\lambda^{(t-k + 1)})
\end{eqnarray*}
If $\overline{LB}_{t-t_W+1} \ge \max(\bar{LB})$ patience = 0, else patience = patience +1

\item If patience $\ge$ P, `stop = TRUE`

\item Set $t:=t+1$
\end{enumerate}

\end{appendix}

%% -----------------------------------------------------------------------------

\end{document}